\documentstyle[prl,aps,twocolumn,epsf]{revtex}
\begin{document}
\twocolumn[\hsize\textwidth\columnwidth\hsize\csname@twocolumnfalse\endcsname

\title{
Beyond the time independent mean field 
theory for nuclear and atomic reactions:\\
Inclusion of particle-hole
correlations
in a generalized random phase approximation}

\author{J.C. Lemm}
\address{
Institut f\"ur Theoretische Physik, Universt\"at M\"unster, 48149 M\"unster,
Germany}
\author{B.G. Giraud}
\address
{Service de Physique Th$\acute{\rm{e}}$orique, DSM--CEA Saclay, 91192
Gif/Yvette, France}
\author
{A. Weiguny\cite{permanent}}
\address
{Yukawa Institute for Theoretical Physics, Kyoto University,
Kyoto 606, Japan}
\date{\today}
\maketitle

\begin{abstract}
The time independent mean field method (TIMF) for scattering
defines biorthonormal sets of single-particle wave functions and
corresponding creation and annihilation operators. $2p - 2h$
correlations can be introduced through a generalized random phase
approximation; $1p - 1h$ contributions vanish
(Brillouin theorem).
While the general variational method for scattering
by Giraud and Nagarajan 
solves inhomogeneous Euler
equations by inversion of the standard, hermitean Hamiltonian,
the present approach diagonalizes a non-hermitean Hamiltonian,
which carries the information about entrance and exit channels.
\end{abstract}

] 
\narrowtext

The time independent mean field theory of collisions (TIMF) \cite{1} has
been tested successfully for solvable cases of 3 and 4 particles \cite{2}
and has been applied to various nuclear and atomic reactions \cite{3}.
It is then natural to go beyond the mean field approach by
introducing particle--hole ({\it ph}) correlations in analogy to the case of bound
states. In this letter we outline how this goal can be reached by a
generalized random phase approximation. Our approach can be viewed as 
an alternative of more traditional inclusions of correlations into scattering
theories, such as the antisymmetrization of cluster structures 
\cite{RGM} and fluctuations 
into time dependent Hartree--Fock \cite{TDHF}, for instance.

We start from a time independent functional \cite{1}
which calculates
the Green function $D(z)$ between incoming and outgoing channel waves
$\chi, \chi'$. The variation of
\begin{equation}
F(\Psi', \Psi) = <\chi'|\Psi> <\Psi'|\chi>/<\Psi'|(z - H)|\Psi>,
\end{equation}
with ${\rm{Im}} \, z > 0$
gives, in appropriate
normalization,
\begin{equation}
(z - H)|\Psi> = |\chi>\, , \quad <\Psi'|(z - H) = <\chi'| \, ,
\end{equation}
\begin{equation}
D(z) = <\chi'|(z - H)^{-1}|\chi> = <\chi'|\Psi> = <\Psi'|\chi> \, ,
\end{equation}
with $\Psi, \Psi'$ as stationary (saddle!) points of (1). For simplicity
we assume that $\chi$ and $\chi'$ are Slater determinants made of
$N$ square integrable orbitals $\chi_{i}$ and $\chi'_{i}$, respectively.
$H$ is the standard full Hamiltonian of $N$ particles. In the same way
one may obtain the $T$--matrix by multiplying $\chi, \chi'$ by prior and
post potentials $V, V'$ in the above equations. While one normally
calculates $D$ as function of (complex) energy $z$, 
{\it we now reverse the strategy}: 
We rewrite (2) with (3) as
\begin{mathletters} \begin{eqnarray}
(z - H)|\Psi> &=& \Lambda |\chi> <\chi'|\Psi>\, ,\\ 
 <\Psi'|(z - H) &=& \Lambda <\Psi'|\chi> <\chi'| \, ,
\end{eqnarray} \end{mathletters} \noindent
which we interpret as right and left eigenvalue equations of the
non--hermitean Hamiltonian
\begin{equation}
H' = H + \Lambda |\chi> <\chi'|\, ; \quad \Lambda = 1/D \,\, \,
{\rm{complex}} \, ,
\end{equation}
with eigenvalues $z = z(\Lambda)$. Choosing fixed values of
$\Lambda$ in (4) and (5) preserves the equivalence of (2) and (4), since
Eqs.(4) are invariant under normalization of $\Psi, \Psi'$ and their
solutions are stationary points of the functional (1) which also is
invariant under normalization of $\Psi, \Psi'$.

The discrete eigenvalues of $H'$ are solutions of the obvious
quantization condition
\begin{equation}
\Lambda^{-1} = <\chi'| (z - H)^{-1}|\chi> \, ,
\end{equation}
which holds for both right and left eigenstates according to Eqs.(4).
It is clear that any real $E$ of the discrete spectrum of $H$ generates
a complex $z$ close to $E$, at least whenever $\Lambda$ is small.
Conversely, there could be discrete eigenvalues $z$ of $H'$ which are
not in the vicinity of discrete eigenvalues of $H$. 
As long as $\Lambda$ is the reciprocal of a physical amplitude $D$,
an additional, discrete, complex eigenvalue emerges,
close to the physical scattering energy $E$.
The continuum of $H'$ is the same as that of $H$ 
because of the compactness \cite{reedsimon}
of  $|\chi> <\chi'|$.

Square integrable left and right eigenstates of 
$H'$ can be found from
the Rayleigh--Ritz--like functional
\begin{equation}
F'(\Psi', \Psi) = <\Psi'| (z - H') |\Psi> \, ,
\end{equation}
where $z$ appears as a norm and phase control Lagrange multiplier,
actually equal to the desired eigenvalue of $H'$, and the trial
functions $\Psi, \Psi'$ are square integrable as long as ${\rm{Im}} \, z
\not=
0$. Indeed, the variational equations of (7) induce the right and left
eigenvalue equations (4). 
It should be stressed here that, from the point
of view of the variational principle for $F'$, one looks for $z$ when
$\Lambda$ is given, while from the point of view of $F$, one looks for
$\Lambda$ when $z$ is given.

In this paper we investigate solutions of (4) where ${\rm{Re}} \, z$ can
be interpreted as an energy of the continuum of $H$, with 
${\rm{Im}} \, z > 0$, 
and we consider $z$ as a function of $\Lambda$. In other words,
we seek an energy $z$ for which the Green function $D(z)$ is equal to
a given value $\Lambda^{-1}$. Our way of solving (6) for $z$ consists
in
diagonalizing $H'$ via suitable approximations of $\Psi, \Psi'$ in the
functional (7). This involves first a Hartree--Fock approximation which
induces a zero order answer $z_{0}(\Lambda)$, then possibly a
Tamm--Dancoff approach (TDA) to the spectrum of $H'$ and finally random
phase (RPA), second order corrections $\Delta z (\Lambda)$, with the
final result $z_{2} (\Lambda) = z_{0} (\Lambda) + \Delta z (\Lambda)$.
First order corrections are automatically cancelled by the Hartree--Fock
method (generalized Brillouin theorem).

The Hartree--Fock approximation
to the diagonalization of $H'$ consists in restricting $\Psi, \Psi'$
to just Slater determinants $\Phi, \Phi'$, made of orbitals
$\varphi_{i}, \varphi'_{i}, i = 1,2,\ldots N$. To simplify the
corresponding variational equations, we redefine $\chi_{i}, \chi'_{i}$
by the conditions
\begin{equation}
<\varphi'_{i}|\chi_{j}> = 0 = <\chi'_{i}|\varphi_{j}> \, \, {\rm{for}}
\,
\, i \not= j \, .
\end{equation}
This just means a linear rearrangement of the
orbitals $\chi_{i}, \chi'_{i}$, leaving the determinants $\chi, \chi'$
invariant. 
Then the term \mbox{$\Lambda <\Psi'|\chi> <\chi'|\Psi>$}
is a product of overlaps
\mbox{$<\varphi'_{i}|\chi_{i}><\chi'_{j}|\varphi_{j}>$} and the
Hartree--Fock equations read
\begin{mathletters}\begin{eqnarray}
(\eta_{i} - h) |\varphi_{i}> &=& \lambda_{i} | \chi_{i}>
<\chi'_{i}|\varphi_{i}>\, ,\\
<\varphi'_{i}| (\eta_{i} - h) &=& \lambda'_{i} <\varphi'_{i} |\chi_{i}>
<\chi'_{i}| \, ,
\end{eqnarray} \end{mathletters} \noindent
$\lambda_{i} = \lambda'_{i}$ reciprocal of the single--particle
Green function,
\begin{equation}
\lambda_{i} 
= \lambda'_{i}=
\left( <\chi'_{i}|(\eta_{i} - h)^{-1}|\chi_{i}>\right)^{-1} \, .
\end{equation}

The operator $h = t + U$ contains a self consistent mean field $U$,
similar to the standard Hartree--Fock potential, and the
propagation energies $\eta_{i}$ are also given by self consistent formulae
similar to those defining Hartree--Fock self energies.
From (9), (10) we immediately read off the Hartree--Fock Hamiltonian
corresponding to $H'$,
\begin{equation}
h' = h + \sum \limits^{N}_{j=1} \, \lambda_{j} |\chi_{j}> <\chi'_{j}| \, ,
\end{equation}
bearing in mind (8). Standard manipulations of (9), together
with (8), show that $\varphi_{i}, \varphi'_{i}$ are biorthogonal.
Finally one can use normalizations such that
\begin{eqnarray}
&&<\varphi'_{i}|\varphi_{j}> = \delta_{ij}\, ; \,
\, <\varphi'_{i}|\chi_{j}> =
<\chi'_{i}|\varphi_{j}> = \gamma_{i} \delta_{ij}  \, ,
\\&&
\gamma_{i} = \frac{<\chi'_{i}|(\eta_{i} - h)^{-1} |\chi_{i}>}{
              \sqrt{<\chi'_{i}|(\eta_{i} - h)^{-2} |\chi_{i}>}} \, ,
\end{eqnarray}
since there are still infinitely many ways to adjust the phases and
norms of $\chi_{i}, \chi'_{i}$ in such a way that $\chi, \chi'$ remain
unchanged. We then extend the diagonalization
of $h'$ beyond the first $N$ right and left eigenstates, and obtain
an infinite, biorthonormal set of orbitals $\varphi_{\alpha},
\varphi'_{\beta}$, {\it{presumably}} complete. This set most
likely includes continuum states, to be normalized accordingly,
hence
\begin{eqnarray}
&<\varphi'_{\beta}|\varphi_{\alpha}> = \delta_{\beta \alpha} \, \,
{\rm{or}} \, \, \delta(\eta_{\beta} - \eta_{\alpha}) ,
&\\&
(\eta_{\alpha} - h') |\varphi_{\alpha}> = 0 \, ,\,\,
<\varphi'_{\beta}| (\eta_{\beta} - h') = 0 \, .
&
\end{eqnarray}
The equivalence of inversion and diagonalization
on the many--particle level, based on the functionals $F(\Psi', \Psi)$
and $F'(\Psi',\Psi)$, is reflected on the single--particle level:
The {\it{homogeneous}} equations (9) for $h'$ are strictly
equivalent to the {\it{inhomogeneous}} TIMF equations,
\begin{equation}
(\eta_{i} - h) |\varphi_{i}> = |\chi_{i}>\, , \, \,
<\varphi'_{i}| (\eta_{i} - h) = <\chi'_{i}| \, ,
\end{equation}
obtained from the functional $F (\Phi', \Phi)$ with
$\Phi, \Phi'$ as Slater determinants.
In particular the eigenvalues of 
$h^\prime (\Lambda = 0) = h$, Eq.(15),
coincide with the poles of $D$, 
where $\Lambda =0$, when calculated from Eqs.(16).

To improve the mean field approach by {\it ph} correlations,
we introduce creation and annihilation operators
$a^{\dagger}_{\alpha}, a_{\alpha}, \alpha = 1,2 \ldots \infty$
for orbitals $\varphi_{\alpha}\, ; \, \, a'^{\dagger}_{\beta},
a'_{\beta},
\beta = 1,2
\ldots \infty$ are the corresponding operators for $\varphi'_{\beta}$.
Finally a set of operators $J^{\dagger}_{i}, J_{i}, i = 1,2 \ldots N$ is
introduced for orbitals $\chi_{i}$, and a similar set $J'^{\dagger}_{i},
J'_{i}$ for orbitals $\chi'_{i}$. A second quantization representation
of
$H'$ is then
\begin{eqnarray}
H' &=&  \sum \limits_{\alpha \beta} 
<\varphi'_{\alpha}|t|\varphi_{\beta}>
a^{\dagger}_{\alpha} a'_{\beta} 
\nonumber\\
&&+ \frac{1}{4} \sum \limits_{\alpha \beta \gamma
\delta} <\varphi'_{\alpha} \varphi'_{\gamma} |v| \varphi_{\beta}
\varphi_{\delta}> a^{\dagger}_{\alpha} a^{\dagger}_{\gamma} a'_{\delta}
a'_{\beta} 
\nonumber\\
&&+ \Lambda J^{\dagger}_{1} J^{\dagger}_{2} \cdots
J^{\dagger}_{N} J'_{N}
\cdots J'_{2} J'_{1} \, .
\end{eqnarray} 
In (17) one set only of operator pairs
$a^{\dagger}_{\alpha}, a'_{\beta}$ appears.
It anticommutes canonically
according to (14). With $|\, 0>$ as fermion vacuum we
have
\begin{equation}
|\Phi> = \prod \limits^{N}_{i=1} \, a^{\dagger}_{i} |0>\, , \, \,
<\Phi'| = <0| \prod \limits^{N}_{i=1} \, a'_{i} ,
\end{equation}
as right and left quasi--particle vacua. Hence a complete set of
{\it ph} operators is given by
\begin{equation}
B^{\dagger}_{mi} = a^{\dagger}_{m} a'_{i}, \, \, B'_{mi} = a'_{m}
a^{\dagger}_{i}\, , 
\end{equation}
for $i = 1,2 \ldots N , \, m=N+1,\ldots ,\infty $,
and the ansatz for boson operators on the RPA level is
\begin{equation}
Q_{\nu} = \sum \limits_{mi} \left(X^{\nu}_{mi} B^{\dagger}_{mi} -
Y^{\nu}_{mi} B'_{mi} \right) \, .
\end{equation}
If one sets $Y^{\nu}_{mi} = 0$ a priori, one falls back to the TDA.

While Wick's theorem is available to calculate
the matrix elements of kinetic energy $T$ and two--particle interaction
$V$, the matrix elements of $\Lambda |\chi> <\chi'|$ are less familiar.
They are simplified, however, in the biorthogonal representation (12),
and read,
\begin{mathletters} 
\begin{eqnarray}
<\Phi'|a^{\dagger}_{i} a'_{m}|\chi> &=& <\Phi'|\chi>
<\varphi'_{m}|\chi_{i}>/<\varphi'_{i}|\chi_{i}>, \nonumber\\
\\
<\chi'|a^{\dagger}_{m} a'_{i}|\Phi> &=& <\chi'|\Phi>
<\chi'_{i}|\varphi_{m}>/<\chi'_{i}|\varphi_{i}>,
\nonumber \\
<\Phi'|a^{\dagger}_{i} a^{\dagger}_{j} a'_{n}
a'_{m}|\chi> &=&
\frac{<\Phi'|\chi>
      <\varphi'_{m} \varphi'_{n}|\chi_{i} \chi_{j}>     } 
     {<\varphi'_{i} |\chi_{i}> <\varphi'_{j} |\chi_{j}> }, \nonumber\\ 
\\
<\chi'|a^{\dagger}_{m} a^{\dagger}_{n} a'_{j}
a'_{i}|\Phi> &=&
\frac {<\chi'|\Phi>
       <\chi'_{i} \chi'_{j}|\varphi_{m} \varphi_{n}>    }
      {<\chi'_{i} |\varphi_{i}> <\chi'_{j} |\varphi_{j}> } \, .
\nonumber
\end{eqnarray}
\end{mathletters} 
Let $M$ be the number of {\it ph} pairs $(m i)$ retained in a
practical calculation. The $2M$ solutions of the RPA equations
\begin{eqnarray}
 <\Phi'|\left[[H',Q_{\nu}], B^{\dagger}_{nj}\right] |\Phi> &=&
E^{*}_{\nu} <\Phi'| [Q_{\nu}, B^{\dagger}_{nj}] |\Phi>
\nonumber
\\
\\ 
 <\Phi'|\left[[H',Q_{\nu}], B'_{nj}\right] |\Phi> &=&
E^{*}_{\nu} <\Phi'| [Q_{\nu}, B'_{nj}] |\Phi>
\nonumber
\end{eqnarray}
then split into 2 families of $M$ solutions.
For the first family, $E^{*}_{\nu}$ denotes the "excitation" energy
which separates the RPA--correlated eigenstate $|\Psi>$ based on $|\Phi>$
from another eigenstate of $H'$ defined by the RPA ansatz $|\Psi_{\nu}>
= Q_{\nu} |\Psi>$. In the second family, for which we will use a label
${\overline{\nu}}, -E^{*}_{\nu}$ is the energy difference between
$<\Psi'_{\overline{\nu}}| = <\Psi'| Q_{\overline{\nu}}$ and the
correlated state
$<\Psi'|$ based on $<\Phi'|$. The solutions $\nu$
of the first family can be told from those $\overline{\nu}$ of
the second family by selecting as $E^{*}_{\nu}$ those RPA
eigenvalues close to the eigenvalues of the TDA, which provides only
$M$ solutions. A straightforward, slightly tedious calculation,
gives for (22)
\begin{mathletters} \begin{eqnarray}
\sum \limits_{mi} \left(C_{nj,mi} X^{\nu}_{mi} +
A_{mi,nj} Y^{\nu}_{mi}\right) &=& -E^{*}_{\nu} Y^{\nu}_{nj}
\\ 
\sum \limits_{mi} \left(A_{nj,mi} X^{\nu}_{mi} + B_{nj,mi}
Y^{\nu}_{mi}\right) &=& +E^{*}_{\nu} X^{\nu}_{nj}
\end{eqnarray} \end{mathletters}
with
\begin{eqnarray}
C_{nj,mi} &=&  -<\varphi'_{i} \varphi'_{j}|v|\varphi_{m}
\varphi_{n}>  
- \bar \Lambda 
\frac {<\chi'_{i}\chi'_{j}|\varphi_{m} \varphi_{n}>    } 
      {<\chi'_{i}|\varphi_{i}> <\chi'_{j}|\varphi_{j}> },
\nonumber \\ \\
 B_{nj,mi} &=& -<\varphi'_{m} \varphi'_{n}|v|\varphi_{i}
\varphi_{j}>
- \bar \Lambda 
\frac{<\varphi'_{m}\varphi'_{n}|\chi_{i} \chi_{j}> }
     {<\varphi'_{i}|\chi_{i}> <\varphi'_{j}|\chi_{j}> },
\nonumber\\
\end{eqnarray}
\begin{eqnarray}
A_{nj,mi} &=& \delta_{mn} \delta_{ij} (\eta_{m} -
\eta_{i}) -
<\varphi'_{n} \varphi'_{i}|v|\varphi_{m} \varphi_{j}>
\nonumber\\
&&+\bar \Lambda \left\{
\frac{ <\chi'_{i}|\varphi_{m}> <\varphi'_{n}|\chi_{j}> } 
     { <\chi'_{i}|\varphi_{i}> <\varphi'_{j}|\chi_{j}> } 
\right.
\nonumber\\
&&-
\left.
\delta_{ij}\sum \limits_{l}
\frac{ <\varphi'_{n}|\chi_{l}> <\chi'_{l}|\varphi_{m}> }
     { <\chi'_{l}|\varphi_{l}> <\varphi'_{l}|\chi_{l}> } \right\}  \, ,
\end{eqnarray} 
\noindent
using $\bar \Lambda =\Lambda <\Phi'|\chi> <\chi'|\Phi> $ 
and the relation
\begin{equation}
\lambda_{i} = \lambda'_{i} = \bar \Lambda  
     /<\varphi'_{i}|\chi_{i}> <\chi'_{i}|\varphi_{i}>  \, .
\end{equation}
Eqs.(23) show less symmetry than the usual RPA equations \cite{4},
as expected from the non--hermitean $H'$ as compared to $H$.
However, $B$ and $C$ are symmetric under exchange of the index 
pairs $(mi)$ and $(nj),$ the block matrix
\begin{equation}
S = \left ( \begin{array}{cc}
C & {\tilde{A}}\\
A & B \end{array} \right)
\end{equation}
is symmetric and
Eqs.(23) read
\begin{equation}
S {X^{\nu} \choose Y^{\nu}} = E^{*}_{\nu} {\cal{M}} {X^{\nu} \choose
Y^{\nu}}, \, {\rm with} \,\,
{\cal{M}} = \left( \begin{array}{cc}
0 & -1\\
1 & 0
\end{array} \right) \, .
\end{equation}
It is then easy to show that the "symplectic diagonalization" problem
(29) is equivalent to diagonalizing the antisymmetric matrix
\begin{equation}
{\cal{A}} = S^{-1/2} {\cal{M}} S^{-1/2} \, ,
\end{equation}
namely
\begin{equation}
{\cal{A}}S^{1/2} {X^{\nu} \choose Y^{\nu}} = E^{* -1}_{\nu} S^{1/2}
{X^{\nu}
\choose Y^{\nu}} \, .
\end{equation}
An elementary derivation of (31) from (29) involves the definition of any
symmetric square root of $S$ and its inversion, and the assumption that
$E^{*}_{\nu} \not= 0$. While translational and rotational
symmetry of $H$ give rise to null eigenvalues in usual $RPA$, the
$|\chi> <\chi'|$ -term in $H'$ destroys these symmetries, hence
eigenvalues $E^{*}_{\nu} = 0$ are usually not to be expected. Special
proofs are available when an exceptional incident occurs.

The fact that ${\cal{A}}$ is antisymmetric proves that the spectrum
splits
into two families of opposite eigenvalues, as announced. Consider now
the $M$ eigenvalues $E^{*}_{0 \nu}$, coming from the diagonalization of
$A$, and define the auxiliary "perturbative RPA" matrix
\begin{equation}
S_{\epsilon} = \left( \begin{array}{cc}
\epsilon C & {\tilde{A}}\\
A & \epsilon B \end{array} \right) \, , \quad 0 \le \epsilon \le 1 \, ,
\end{equation}
with corresponding ${\cal{A}}_{\epsilon}$. It is clear that, as
$\epsilon \rightarrow 0$, half of the eigenvalues of the auxiliary
problem (32) converge towards $E^{*}_{0 \nu}$, while the other $M$
eigenvalues tend to $-E^{*}_{0 \nu}$. By continuity when $\epsilon
\rightarrow 1$, this makes it easy to separate the eigenvalues of (29)
into the two announced families, both being complex due to the
nonhermiticity of $H'$. With the first family, the usual bosonic
interpretation of the RPA--like Hamiltonian finally provides the
correction
\begin{equation}
\Delta z (\Lambda) = \frac{1}{2} \left( \sum \limits^{M}_{\nu = 1} \,
E^{*}_{\nu} - {\rm{Tr}} \, A \right) \, .
\end{equation}
A function inversion of the final result
$
z_{2} (\Lambda) = z_{0} (\Lambda) + \Delta z (\Lambda) \, ,
$
generating a function $\Lambda (z_{2})$, then provides the improvement
on the Green function, beyond the mean field, by inclusion
of {\it ph} correlations.

In conclusion, the key element of our study is the
equiv\-alence of the following two problems: \\
1.) Calculate the Green function $D(z)$ at some fixed (complex) energy
$z$ by operator inversion of the exact Hamiltonian $H$.
One solves an {\it{inhomogeneous}} problem with channel wave functions
$\chi, \chi'$ as inhomogeneity.\\
2.) Diagonalize a Hamiltonian $H'$ which differs from $H$ by the
operator $\Lambda |\chi> <\chi'|$, comprising the reaction boundary
conditions. In this {\it{homogeneous}} approach, one calculates the
(complex) eigenvalues $z$ as function of the parameter $\Lambda = 1/D$
and obtains the desired Green function by inversion of the function
$z(\Lambda)$.\\
Thus operator inversion is replaced by operator diagonalization followed
by function inversion. While problems 1.) and 2.) can equally well be
solved in the mean field approximation, we use 2.) to incorporate
correlations in the wave function in terms of {\it ph} bosons.
Although the resulting boson Hamiltonian is less symmetric than
in standard (hermitean) RPA, the remaining symmetry is sufficient to
diagonalize the (non--hermitean) boson Hamiltonian by canonical, but
non--unitary transformation and to calculate $z(\Lambda)$ in this boson
approximation.

We tested numerically the above method for a
one-dimensional, separable, two-body Hamiltonian
of relative motion
\begin{eqnarray}
\lefteqn{<q^\prime Q^\prime | {\cal H} |q Q >= }
\label{potrelativ}
\\&&
\delta (Q-Q^\prime)
\left[
q^2 \delta (q-q^\prime) -\lambda \nu^2 qq^\prime
\exp (-\nu^2 q^2)\exp (-\nu^{\prime 2} q^{\prime 2})
\right] ,
\nonumber
\end{eqnarray}
with $q$, $Q$ the relative and total momentum,
which gives an obvious solution 
\cite{3} for the exact amplitude $D(z)$ of (3). Our TIMF and RPA
numerics, however, used the single-particle representation 
which becomes mandatory for antisymmetrized, many-particle calculations.
The channel waves $\chi$, $\chi^\prime$ were taken as boosted Gaussian
functions.
Mass and potential parameters as well as boost and width of $\chi,\chi^\prime$
were chosen to be typical for nuclear physics problems;
for technical details we refer to \cite{5}.
The RPA approximation of $D(z)$, based on (23) and (33),
indeed improved the TIMF results (see Fig.1)
obtained by solving (16).

One of us (A. W.) wants to thank the
Japan Society for Promotion of Science and the Deutsche Akademische
Austauschdienst for financial support during his stay in Japan. He is
also very grateful for the hospitality found at the Yukawa Institute for
Theoretical Physics in Kyoto.

\begin{figure}[htb]
\centering
 \mbox{\epsfysize=60mm%
       \epsffile{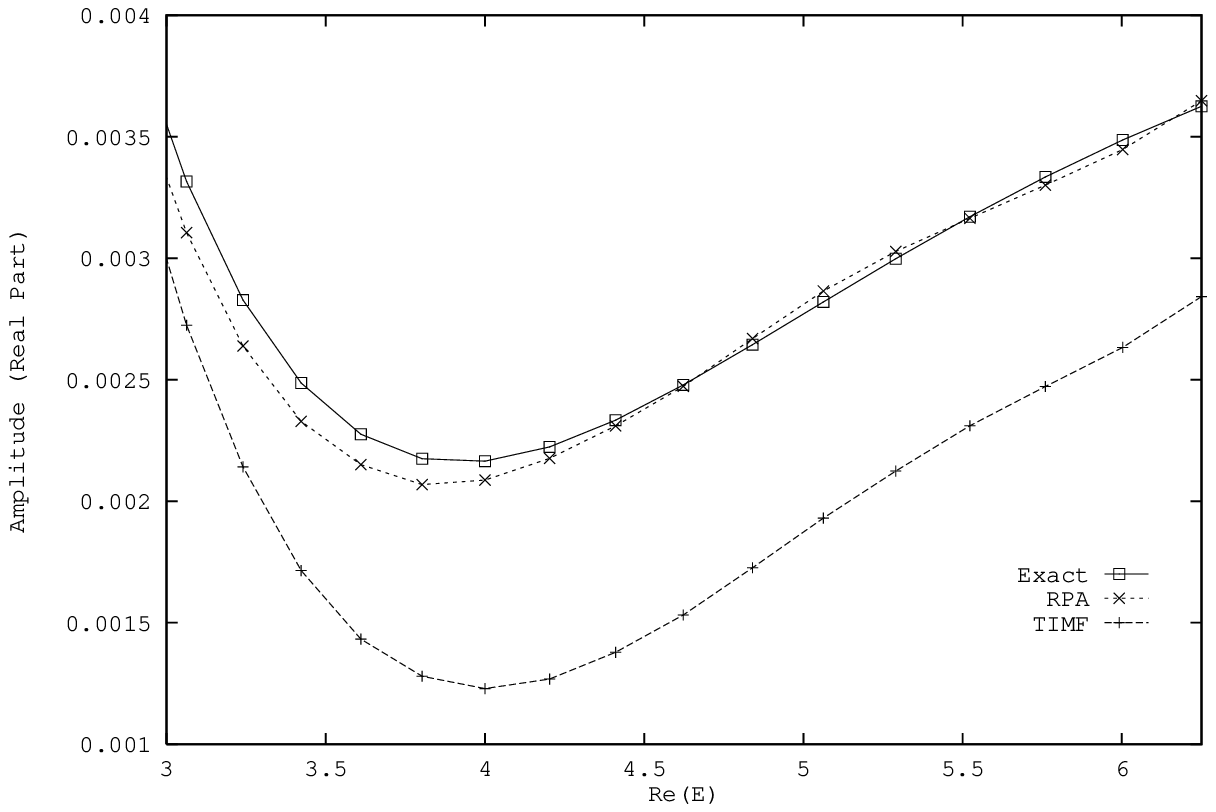}
}
 \mbox{\epsfysize=60mm%
       \epsffile{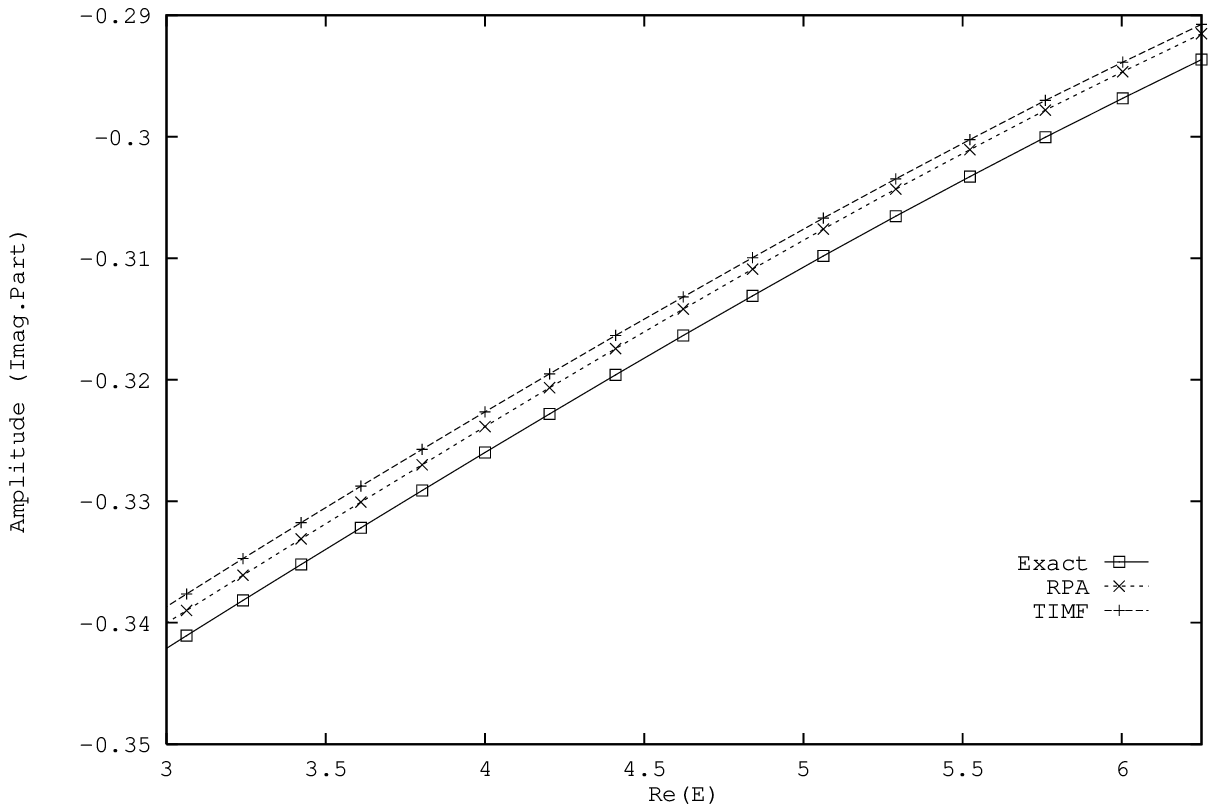}
}
\caption{Real and imaginary part of the exact, 
TIMF-- and RPA--amplitudes vs.\ Re $E$,
at Im $E$=2.0 
for a solvable two--body model with gaussian interaction.}
\label{Klineausr}
\end{figure}


\begin{references}
\bibitem[*]{permanent}
Permanent Address: Institut f\"ur Theoretische Physik, Universit\"at
M\"unster, 48149 M\"unster, Germany.
\bibitem{1} B.G.\ Giraud, M.A.\ Nagarajan, and I.J.\ Thompson, Ann.Phys.(N.Y.)
{\bf 152}, 475 (1984).
\bibitem{2}  Y.\ Abe and B.G.\ Giraud, Nucl.Phys. {\bf A440}, 311 (1985);
J.\ Lemm, A.\ Weiguny, and B.G.\ Giraud, Z.Phys. {\bf A336}, 179
(1990).
\bibitem{3} B.G.\ Giraud, M.A.\ Nagarajan, and C.J.\ Noble, Phys.Rev.A
{\bf 34}, 1034(1986); B.G.\ Giraud and M.A.\ Nagarajan, Ann.Phys.(N.Y.)
{\bf 212}, 260(1991); B.G.\ Giraud, Y.\ Hahn, F.\
Mekideche, and J.\ Pascale, Z.Phys. {\bf D27}, 295(1993).
\bibitem{RGM} Y.C. Tang: {\it Microscopic Description of the Nuclear Cluster
Theory}, Lecture Notes in Physics, Vol.\ 145 (Springer, New York, 1981).
\bibitem{TDHF} R.\ Balian and M.\ V\'en\'eroni, Ann.Phys.(N.Y.)
{\bf 195} (1989), 324.
\bibitem{reedsimon} M.\ Reed and B.\ Simon: {\it Methods of Modern Mathematical
Physics. I: Functional Analysis} (Academic Press, New York, 1972).
\bibitem{4} D.J.\ Thouless, {\it The Quantum Mechanics of Many Body Systems}
(Academic Press, New York, 1961).
\bibitem{5} J.C.\ Lemm, doctoral dissertation, M\"unster University, 1993.
\end{references}
\end{document}